\documentclass[useAMS,usenatbib]{mn2e}
\usepackage{amssymb}
\usepackage{amsmath}
\usepackage{graphicx}
\usepackage{rotating}
\usepackage{lscape}
\usepackage{array}
\usepackage{wrapfig}
\usepackage{graphics}
\usepackage{hyperref}
\usepackage{color}

\usepackage{subfig}
\def\hi{H\,{\sc i}}
\def\hii{H\,{\sc i}{\sc i}}

\def\kms{km~s$^{-1}$}
\def\msun{M$_{\odot}$}
\def\Ha{H$\sf \alpha$}
\def\micron{$\mu$m}

\def\msunpps{M$_{\odot}$~pc$^{-2}$}

\long\def\symbolfootnote[#1]#2{\begingroup%
\def\thefootnote{\fnsymbol{footnote}}\footnote[#1]{#2}\endgroup} 
\def\aj{AJ}%
\def\araa{ARA\&A}%
\def\apj{ApJ}%
\def\apjl{ApJ}%
\def\apjs{ApJS}%
\def\aap{A\&A}%
\def\aapr{A\&A~Rev.}%
\def\mnras{MNRAS}%
\title[An HI study of NGC~3521]{An \hi\ study of NGC~3521~-~a galaxy with a slow-rotating halo}
\author[E.~C.~Elson]{E.~C.~Elson$^{1}$\thanks{E-mail:
elson.e.c@gmail.com (ECE)}\\
$^{1}$Astrophysics, Cosmology and Gravity Centre (ACGC), Department of Astronomy, University of Cape Town,\\ Private Bag X3,
Rondebosch 7701, South Africa}

\begin{document}

\pagerange{\pageref{firstpage}--\pageref{lastpage}} \pubyear{2013}

\maketitle

\label{firstpage}

\begin{abstract}

A study is presented of \hi\ line observations of the nearby spiral galaxy NGC~3521 observed with the VLA as part of The \hi\ Nearby Galaxy Survey.  Clearly evident in the \hi\ data cube is the presence of an anomalous \hi\ component that is both diffuse and slow-rotating.  The data cube is dynamically decomposed into regular and anomalous \hi\ components.  A mass of $\mathrm{M_{HI}}=1.5\times 10^9$~\msun\ is estimated for the anomalous \hi\ - 20~per~cent of the total \hi\ mass.  Standard \hi\ data products and rotation curves are produced for each dynamical component.  In terms of circular rotation speed, the anomalous \hi\  is found to lag the regular \hi\  by $\sim 25$~-~125~\kms.  Three-dimensional models are generated and used to determine the possible location of the anomalous \hi.  The results strongly suggest it to be distributed in a thick disc with a scale-height of  a few kpc ($\sim 3.5$~kpc).  It is concluded that the anomalous \hi\ in NGC~3521 constitutes a slow-rotating halo gas component, consistent with similar findings for other nearby galaxies.  {\color{black}A study of the radial distribution of the anomalous \hi\ shows it to be spatially coincident with the inner regions of the stellar disc where the star formation rate is highest.}  It is most likely a galactic fountain that has deposited gas from the disc of the galaxy into the halo.

\end{abstract}


%INTRODUCTION----------------------------------------------------------------------------------------------------------------------------------------------------------------------------------------------
\section{Introduction}\label{sec_intro}

The cycle of gas in galaxies is known to play an important role in their evolution.  Over the years a general picture has developed of gas being contained not only within the discs of galaxies, but also in their halos.  The mechanisms by which  gas arrives in the halo have been the focus of many studies.  \citet{galactic_fountain} first proposed the ``galactic fountain'' scenario in which gas is ejected by star formation from the disc of a galaxy in its halo.  The gas expands and cools in the halo, subsequently raining back down onto the disc.  Observations of nearby galaxies such as NGC~891 \citep{oosterloo_NGC891} and NGC 6946 \citep{NGC6946_kamphius} show their extra-planar gas to be concentrated very close to the star-forming disc, suggesting it to indeed be driven by star formation.  

Halo gas can also be of  external origin.  {\color{black}Simulations show the majority of baryons to enter the virial radius of a galactic halo in gaseous form along cosmic filaments (\citealt{putman_2012_review} and references therein).  Cold mode accretion is found to dominate in galaxies at high redshifts ($z\gtrsim 2$), as well as in present day low-mass ($M_{halo}\lesssim 10^{12}$~\msun) galaxies.}  Several bodies of observational evidence exist for the prevalence of gas accretion.  The high-velocity clouds seen in the Milky Way halo are now known to be infalling intergalactic gas \citep{wakker_2007, wakker_2008}.  Their low metallicities  point directly to an external origin {\color{black}- either the intergalactic medium or accreted low-mass satellites.}  Deep observations of nearby galaxies reveal the presence of extended \hi\ filaments with peculiar velocities.  These extended gas structures are often characterised by rotation-speed gradients in the direction perpendicular to the disc of the galaxy (e.g.~\citealt{oosterloo_NGC891}) as well as global inflow motion (e.g.~\citealt{ngc2403_beard}).  Observational estimates of the accretion rate vary from $\sim 0.1$~-~0.2~\msun~yr$^{-1}$ to values possibly ten times higher, as needed to replenish the gas depleted by star formation (\citealt{sancisi_review} and references therein).  

The halo gas in galaxies seen edge-on is oftentimes directly observable (e.g. NGC~891).  {\color{black}However, for galaxies seen at more intermediate inclinations the presence of extra-planar gas needs to be inferred very carefully from the observed kinematics (e.g.~\citealt{boomsma_6946}~-~NGC~6946, \citealt{ngc2403_beard}~-~NGC~2403, \citealt{hess_2009}~-~NGC~2997).}  This work focuses on an \hi\ dynamical study of one such galaxy, NGC~3521.

At a distance of 10.7~Mpc\symbolfootnote[2]{Calculated from NED assuming Hubble flow distance corrected for Virgo infall.} \citep{THINGS_deblok} NGC~3521 is a late-type galaxy that has been studied at various wavelengths by many investigators.  The system has a star formation rate of 2.1~\msun~yr$^{-1}$ \citep{leroy_THINGS} and an integrated \Ha\ luminosity L$_{\mathrm{H\alpha}}=10^{32}$~J~s$^{-1}$ \citep{kennicutt_SF_2008}.  \citet{trachternach_THINGS} estimate the dynamical centre of the galaxy to be $\alpha_{2000}$~=~11$^{\mathrm{h}}$~05$^{\mathrm{m}}$~48.$^{\mathrm{s}}$6, $\delta_{2000}$~=~$-00^{\circ}$~02$'$~09$''$.  This closely matches the position of the photometric centre, and is adopted as the centre position of the galaxy throughout this work.  \citet{regan_2006} use IRAC imaging to estimate the stellar disc scale-length of NGC~3521 to be $1.7\pm 0.2$~kpc, and trace the stellar exponential disc out to about 8~-~9 disc scale-lengths.  The total stellar mass of the galaxy is estimated to be $\mathrm{M_*}= 5\times 10^{10}$~\msun\ \citep{leroy_THINGS}.  Table~\ref{properties_table} summarises the main properties of NGC~3521.

NGC~3521 is a gas-rich system with a total neutral gas mass $\mathrm{M_{HI}}=8.02\times 10^9$~\msun\symbolfootnote[3]{For a distance of 10.7~Mpc.} \citep{THINGS_walter}.  It has been studied in \hi\ line emission by several authors.    \citet{NGC3521_casertano} {\color{black}used} VLA observations of the galaxy to derive a rotation curve characterised by a large decrease in rotation velocity ($\gtrsim 50$~\kms) between 1 and 3 optical radii.  More recently, the system was re-observed with the VLA as part of The \hi\ Nearby Galaxy Survey (THINGS, \citealt{THINGS_walter}).  \citet{THINGS_deblok} produce a rotation curve for the galaxy that does not drop significantly in the outer parts.  {\color{black}A straight line fitted to the ``flat'' portion of their rotation curve (from the knee to the last measured point) drops from 234.7~\kms\ at $R=120$~arcsec to 186.0~\kms\ at $R=600$~arcsec.  Based on the last measured point on the rotation curve, the dynamical mass of NGC~3521 is $\mathrm{M_{dyn}}\sim 3\times 10^{11}$~\msun.}  

%Using an NFW halo parameterisation of the dark matter distribution in NGC~3521, \citet{THINGS_deblok} 

%\begin{table}
%\begin{center}
%\caption{Summary of NGC~3521 properties.}
%\label{properties_table}
%\begin{tabular}{cccc}
%\hline
%\hline
%\\
%&Parameter			&		Value		&		Reference\\
%\\
%\hline
%\\
%1&Distance [Mpc]		&		10.7		&		[1]			\\
%2&R.A. (J2000) 	&	11$^{\mathrm{h}}$~05$^{\mathrm{m}}$~48.$^{\mathrm{s}}$6	&	[2]\\		
%3&Decl. (J2000) 	&	$-00^{\circ}$~02$'$~09$''$	&	[2]\\		
%4&$\log_{10}\mathrm{SFR}$ [\msun~yr$^{-1}$]		& 	3.3			&	[3]\\
%5&$\log_{10}\mathrm{L_{H\alpha}}$ [J~s$^{-1}$]		&	32.0			&	[4]	\\
%6&$\alpha_{3.6}$ [kpc]							&	1.7				&	[5]				\\
%7&$\log_{10}$M$_*$ [\msun]						&	10.7			&	[6]			\\
%8&$\log_{10}\mathrm{M_{HI}}$ [\msun]			&	9.9				&	[7]		\\
%9&$\mathrm{V_{rad}}$ [\kms]						&	798.2			&	[7]		\\
%10&$\mathrm{M_{dyn}}$ [\msun]					&	$3\times 10^{11}$	&	[1]	\\
%\\
%\hline
%\end{tabular}
%\end{center}
%Note. -- (1) heliocentric distance; (2/3) dynamical centre right ascension/declination; (4) \Ha\ star formation rate; (5) integrated \Ha\ luminosity; (6) 3.6~\micron\ disc scale length; (7) stellar mass; (8) \hi\ mass, (9) heliocentric radial velocity, (10) dynamical mass.\\
%\\
%References. -- [1] \citealt{THINGS_deblok}; [2] \citealt{trachternach_THINGS}; [3]~\citealt{lee_2009}; [4]~\citealt{kennicutt_SF_2008}; [5]~\citet{regan_2006}; [6]~\citealt{leroy_THINGS}; [7]~\citealt{THINGS_walter}.
%\end{table}%
 
 \begin{table*}
\begin{center}
\caption{Summary of NGC~3521 properties.}
\label{properties_table}
\begin{tabular}{ccc}
\hline
\hline
\\
			Parameter			&		Value		&		Reference\\
\\
\hline
\\
Heliocentric distance [Mpc] 						&	10.7					&		 \citealt{THINGS_deblok}		\\
Right Ascension (J2000) 								&	11$^{\mathrm{h}}$~05$^{\mathrm{m}}$~48.$^{\mathrm{s}}$6	&	\citealt{trachternach_THINGS}	\\
Declination (J2000) 								&	$-00^{\circ}$~02$'$~09$''$	&		\citealt{trachternach_THINGS}		\\
Star formation rate [\msun~yr$^{-1}$]			& 	2.1			 			&		\citealt{leroy_THINGS}				\\
\Ha\ luminosity [J~s$^{-1}$] 					&	$10^{32.0}$				&		\citealt{kennicutt_SF_2008}		\\
3.6~\micron\ disc scale length [kpc]				&	1.7 						&		\citet{regan_2006}				\\
Stellar mass [$10^{10}$~\msun]				&	5.0						&		\citealt{leroy_THINGS}			\\
\hi\ mass [$10^{9}$~\msun]					&	8.02						&		\citealt{THINGS_walter}			\\
Heliocentric radial velocity [\kms]				& 	798.2					&		\citealt{THINGS_walter}			\\
%Dynamical mass							&	$10^{11.4}$~\msun			&		\citealt{THINGS_deblok}			\\
\end{tabular}
\end{center}
\end{table*}%

For each of the \citet{NGC3521_casertano} and \citet{THINGS_deblok} papers, position-velocity slices extracted along the \hi\ major axis of the galaxy reveal clear evidence for asymmetric and/or double-component \hi\ line profiles.  The line profiles are skewed towards the systemic velocity of the galaxy and are manifested in the position-velocity slices as an ``\hi\ beard''.  Other galaxies have been observed to have \hi\ beards, for example: NGC~2403 \citep{Schaap_2000, ngc2403_beard}.  The general explanation for such a pattern is the presence of \hi\ gas located above the plane of the galaxy (in a thick disc or in the halo) and rotating more slowly than the disc \citep{Swaters_NGC891}. 

This paper presents a dynamical study of the THINGS \hi\ line observations of NGC~3521.  The goal is to separate the beard emission from the disc emission and to quantify its dynamics as well as its distribution within the galaxy.  

The outline of this paper is as follows.  The \hi\ data are presented and described in Section~\ref{sec_HI_data}.  Section~\ref{sec_profile_parameterisations} describes the method used to decompose the data into two subsets representing the regular and anomalous \hi\ mass components within the galaxy.  Various \hi\ data products are presented in Section~\ref{sec_HI_products}.  Section~\ref{sec_modelling} focuses on modelling the velocity fields of the regular and anomalous \hi\ components in order to generate separate rotation curves.  Full three-dimensional models of NGC~3521 are also presented in Section~\ref{sec_modelling}.  A discussion of the origin of the anomalous \hi\ appears in Section~\ref{anomHI_origin}.  A discussion of the results is given in Section~\ref{sec_discussion}.  The summary and  conclusions are presented in Section~\ref{sec_summary}. 

%HI DATA----------------------------------------------------------------------------------------------------------------------------------------------------------------------------------------------
\section{\hi\ data}\label{sec_HI_data}

In this work the THINGS naturally-weighted (NA) \hi\ data cube of NGC~3521 is utilised; {\color{black}downloaded from the THINGS public data repository}\symbolfootnote[4]{\url{http://www.mpia-hd.mpg.de/THINGS/Data.html}}.  The data were obtained by \citet{THINGS_walter} using the Very Large Array in its B, C, DnC and D configurations for a total of $\sim21$ hours.  The NA cube has a spatial resolution of 14.14~arcsec~$\times$~11.15~arcsec and a channel width of 5.19~\kms.  The reader is referred to Table~2 in \citet{THINGS_walter} for a full description of the observing setups.  

The noise in a line-free channel of the data cube is Gaussian distributed.  For this work, robust statistical methods were used to estimate a mean noise level of -2.1~$\mu$Jy~beam$^{-1}$ and a standard deviation of 3.9~mJy~beam$^{-1}$.  For emission filling the synthesised beam, a flux density of 1.0~mJy~beam$^{-1}$ corresponds to an equivalent \hi\ column density of $\sim3.6\times 10^{19}$~cm$^{-2}$, or an \hi\ mass surface density of $\sim 0.3$~\msunpps\ at the adopted distance of 10.7~Mpc.

%LINE PARAMETERISATIONS----------------------------------------------------------------------------------------------------------------------------------------------------------------------------------------------
\section{Dynamical decomposition}\label{sec_profile_parameterisations}
\begin{figure}
\centering
\includegraphics[width=1.\columnwidth, angle=0]{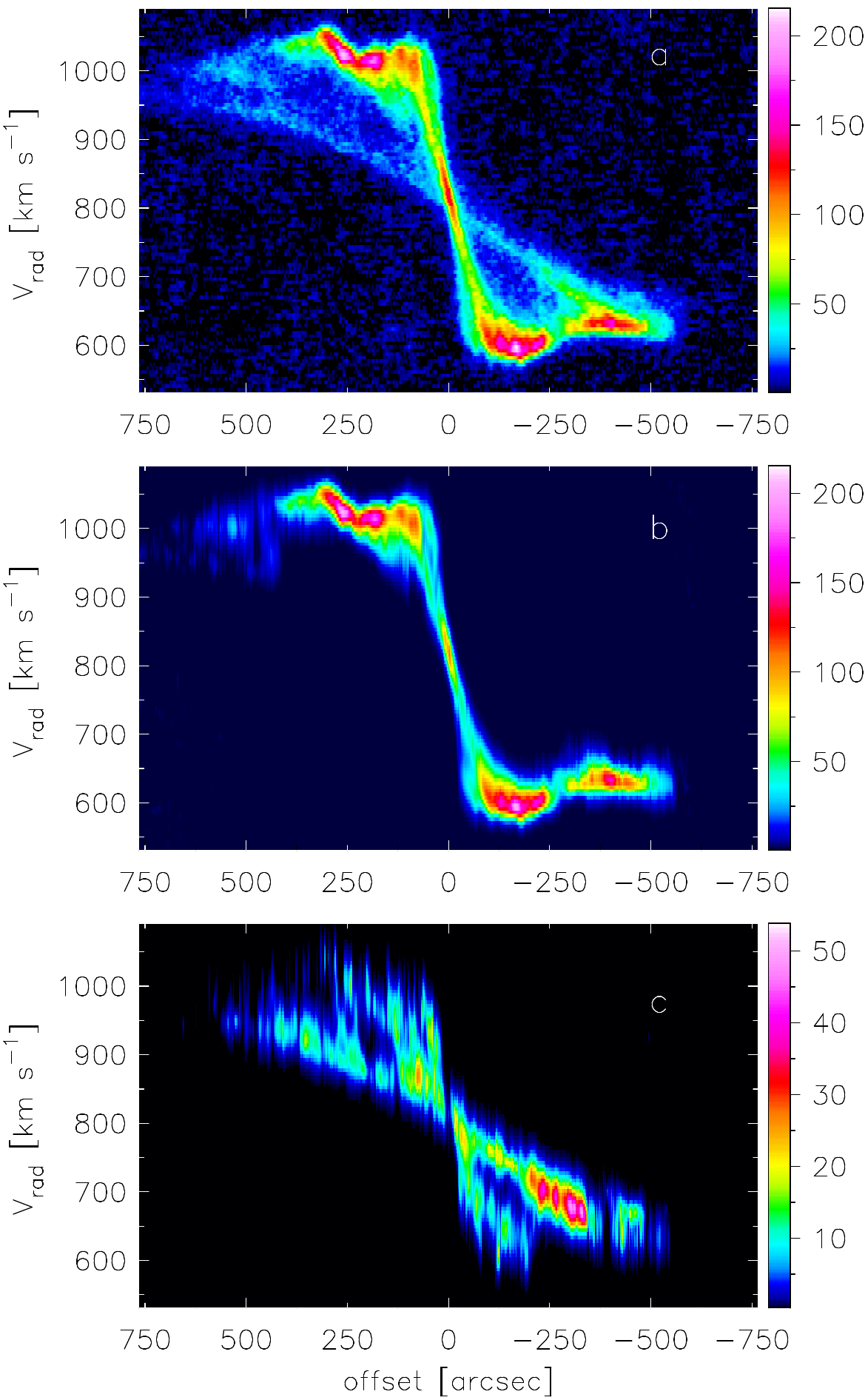}
\caption{Integrated (35-arcsec-thick) major-axis position-velocity slices extracted from the NGC~3521 \hi\ data cubes.  Panel~\textit{a}: the full THINGS cube containing all of the \hi\ emission.  Panel~\textit{b}: the regular \hi\ data cube containing emission from the thin \hi\ disc.  Panel~\textit{c}: the anomalous \hi\ data cube containing  emission from the slow-rotating \hi\ mass component. Identical colour scales are used for panels $a$ and $b$ while a separate scale is used for panel~$c$.  \hi\ column densities to which the colours in each panel correspond are given by the colour bars in units of $10^{20}$~cm$^{-2}$.  Note the much lower column densities of the anomalous \hi\ emission in panel $c$.}
\label{regHI_anomHI_cubes}
\end{figure}

NGC~3521 consists of two dynamically distinct \hi\ components.  Figure~\ref{regHI_anomHI_cubes}a shows an integrated position-velocity slice extracted from a 37.5-arcsec-thick strip placed along the major axis of the galaxy.  This position-velocity slice clearly highlights the bimodal nature of the \hi\ line profiles.  They generally consist of:
\begin{enumerate}
\item A high-intensity, fast-rotating component.  The line-of-sight velocities of this component rise sharply as $V_{rad}\propto R$ for $R\lesssim 50$~arcsec and then remain approximately constant for larger radii.  This is the thin \hi\ disc of the galaxy.
\item A  diffuse (low-intensity) component which at any radius in the position-velocity slice is always located closer to the systemic velocity of the galaxy.  This anomalous \hi\ component is therefore slow-rotating - perhaps distributed outside of the thin \hi\ disc.
\end{enumerate}
The two components are either fully separated in velocity or merge together to form a combined line profile skewed towards the systemic velocity of the galaxy.  The main aim of this work is to decompose the \hi\ line profiles in order to produce:
\begin{enumerate}
\item A cube containing the emission from the thin \hi\ disc, henceforth referred to as the ``regular'' \hi\ cube.
\item A cube containing the diffuse, slow-rotating \hi\ emission, henceforth referred to as the ``anomalous'' \hi\ cube.
\end{enumerate}
The emission in each cube is studied and modelled separately in order to quantify its distribution and kinematics.  

\begin{figure}
\centering
\includegraphics[width=1.\columnwidth, angle=0]{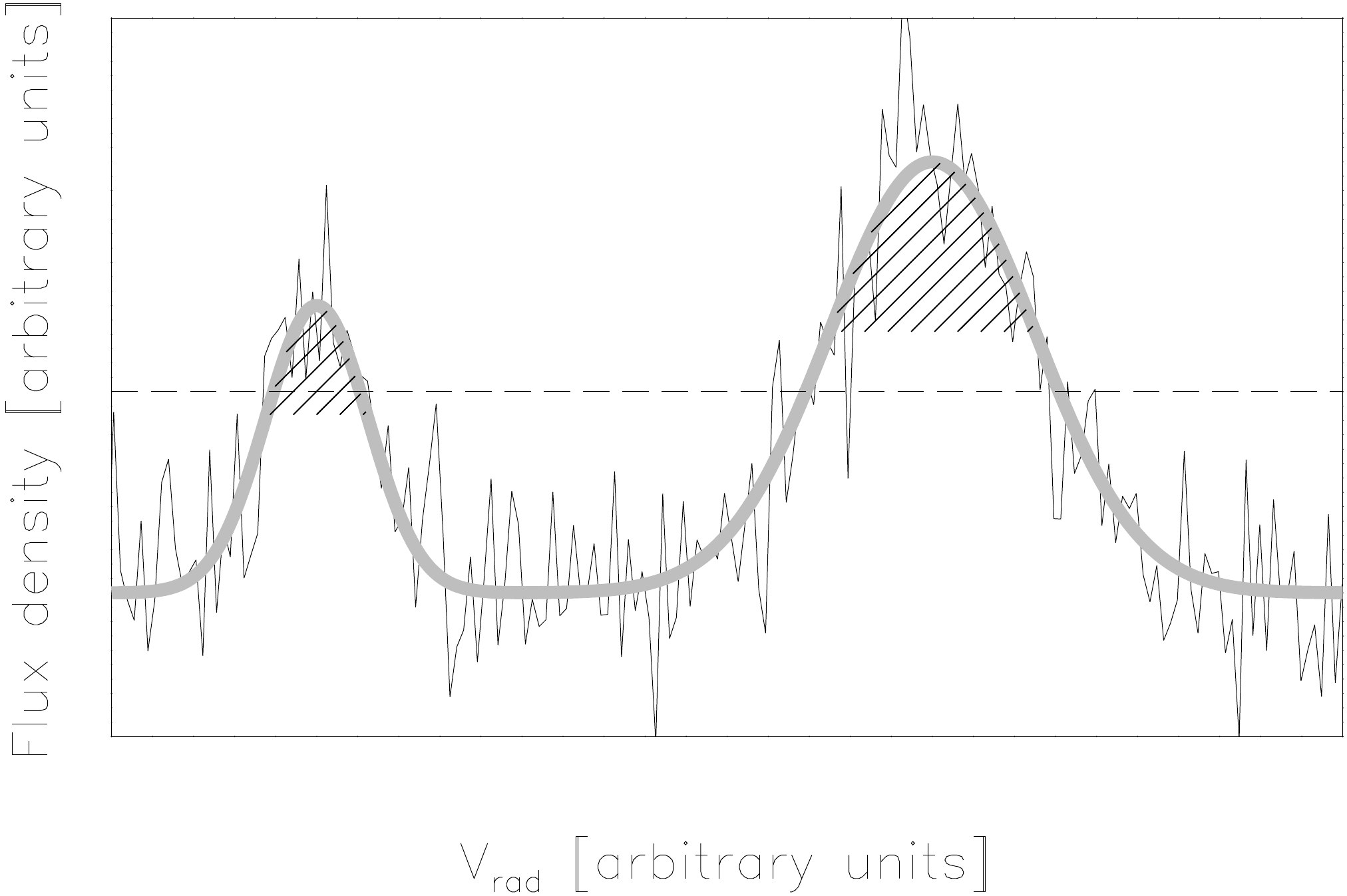}
\caption{{\color{black}Data representing a typical \hi\ line profile (black), fitted by a double Gaussian (grey).  The dashed horizontal line marks the $2\sigma$ noise level of the data.  The hatched region of each Gaussian component represents 20~per~cent of its total area, measured vertically from the peak of the profile towards zero.  The fitted component on the right is accepted, while the component on the left is rejected.}}
\label{illustration}
\end{figure}

{\color{black}Each line profile in the \hi\ data cube was parameterised as a double Gaussian.  {\color{black}In order to avoid the decomposition results being affected by low-signal-to-noise \hi\ emission, only those fitted Gaussian components with more than 20~per~cent of their integrated flux above the $2\sigma$ level of the noise in a line-free channel of the cube were considered ($2\sigma=7.8$~mJy~beam$^{-1}$).  This criterion is henceforth referred to as the ``line-strength'' criterion for fitted profiles.  To illustrate the implementation of the line-strength criterion, Fig.~\ref{illustration} shows a double Gaussian fit to some data representing a typical line profile.  The dashed horizontal line marks the 2$\sigma$ noise level of the data.  The blue-hatched region of each fitted Gaussian component represents 20~per~cent of its total area, measured in the vertical direction from the peak of the profile towards zero.  Because the rightmost component has its entire hatched region above the $2\sigma$ noise level, it satisfies the line-strength criterion and is placed into either the regular or anomalous \hi\ cube.  However, the component on the left does not satisfy the line-strength criterion due to some of its hatched region lying below the dashed line, and is therefore placed into neither cube.}}

\begin{figure}
\centering
\includegraphics[width=1.\columnwidth, angle=0]{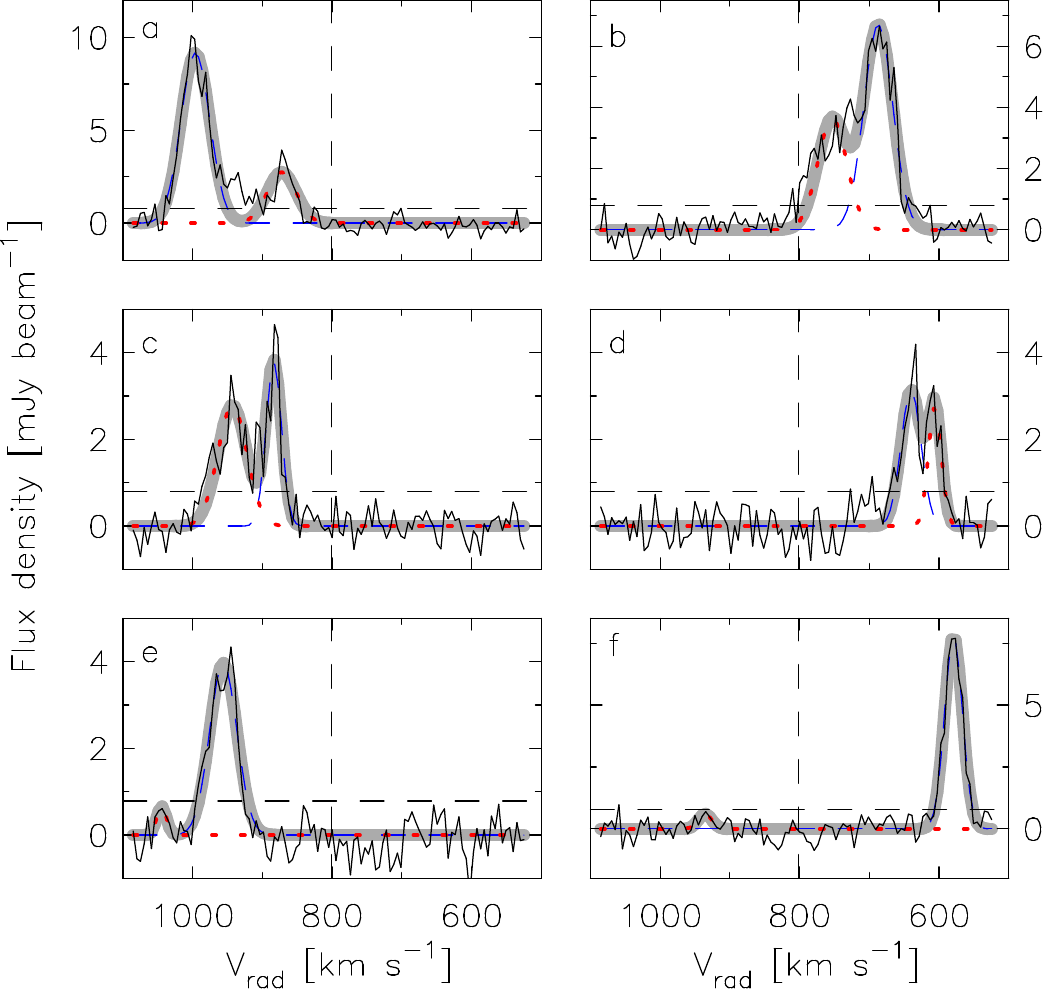}
\caption{Line profiles extracted from the THINGS \hi\ data cube.  In each panel the solid black curve represents the line profile.  The thick grey curve represents the double-component Gaussian fit.  The blue-dashed and red-dotted curves represent the Gaussian components corresponding to the regular and anomalous \hi, respectively.  The vertical and horizontal dashed lines mark the systemic velocity of the galaxy and the $2\sigma$ noise level (7.8~mJy~beam$^{-1}$) of the noise in a line-free channel of the cube, respectively.}
\label{line_profiles}
\end{figure}

{\color{black}The fitted components of the \hi\ line profiles were placed into the regular or anomalous \hi\ cubes, or entirely rejected, in the following ways:}
\begin{enumerate}
\item {\color{black}A line profile with both of its fitted components satisfying the line-strength criterion had its lower-amplitude component placed into the anomalous \hi\ cube if the peak of that component was located closest to the systemic velocity of the galaxy.  The other fitted component was placed into the regular \hi\ cube.  Figures~\ref{line_profiles}a,~b show examples of such line profiles.\\

\item A line profile that has both of its fitted components satisfying the line-strength criterion, yet has its larger-amplitude component lying closer to the systemic velocity than the lower-amplitude component, is inconsistent with the general characteristics of the regular and anomalous \hi\ kinematics as shown in Fig.~\ref{regHI_anomHI_cubes}a.  This figure clearly shows that, in general, it should be the lower-amplitude component (anomalous \hi) that lies closest to the systemic velocity while the larger-amplitude component (regular \hi) is further removed.  Such line profiles (Figs.~\ref{line_profiles}c,~d) were entirely rejected. \\

\item In the case that a line profile had only one fitted component satisfying the line-strength criterion, that component was placed into the regular \hi\ cube.  This is based on the assumption that the regular \hi\ is more intense than the anomalous \hi.  Figures~\ref{line_profiles}e, f show some examples of such line profiles.}
\end{enumerate}

Major-axis position velocity slices extracted from the final versions of regular and anomalous \hi\ data cubes are shown in Figs.~\ref{regHI_anomHI_cubes}b,~c.  These slices show the regular and anomalous \hi\ components to be neatly separated from one another.  The structure of the slice from the regular cube is typical of a circularly-rotating, thin \hi\ disc.  The slice extracted from the anomalous \hi\ cube is typically beard-like, suggesting an extra-planar \hi\ distribution.

%HI MAPS----------------------------------------------------------------------------------------------------------------------------------------------------------------------------------------------
\section{\hi\ data products}\label{sec_HI_products}
Having decomposed the THINGS \hi\ data cube into separate subsets containing the emission from the regular and anomalous \hi\ mass components in NGC~3521, this section presents a set of standard \hi\ data products for each data cube.  

\subsection{Channel maps}
Channel maps of the THINGS \hi\ data cube are shown in Fig.~\ref{channel_maps1}.  In each channel the regular and anomalous \hi\ is roughly delimited by blue-dashed and red-dotted contours, respectively\symbolfootnote[3]{Throughout this work, blue and red curves are used in figures to represent regular and anomalous \hi\ emission, respectively.}.  In order to generate the contours, the regular and anomalous \hi\ data cubes were smoothed down to a spatial resolution of $\sim~45$~arcsec.  In these smoothed cubes the blue-dashed and red-dotted contours represent \hi\ column density levels of $\sim 9\times 10^{18}$ and $3\times 10^{18}$~cm$^{-2}$, respectively.  In the channel maps, the contours clearly show the anomalous \hi\ to dynamically lag the regular \hi.  

\begin{figure*}
\centering
\includegraphics[width=1.8\columnwidth, angle=0]{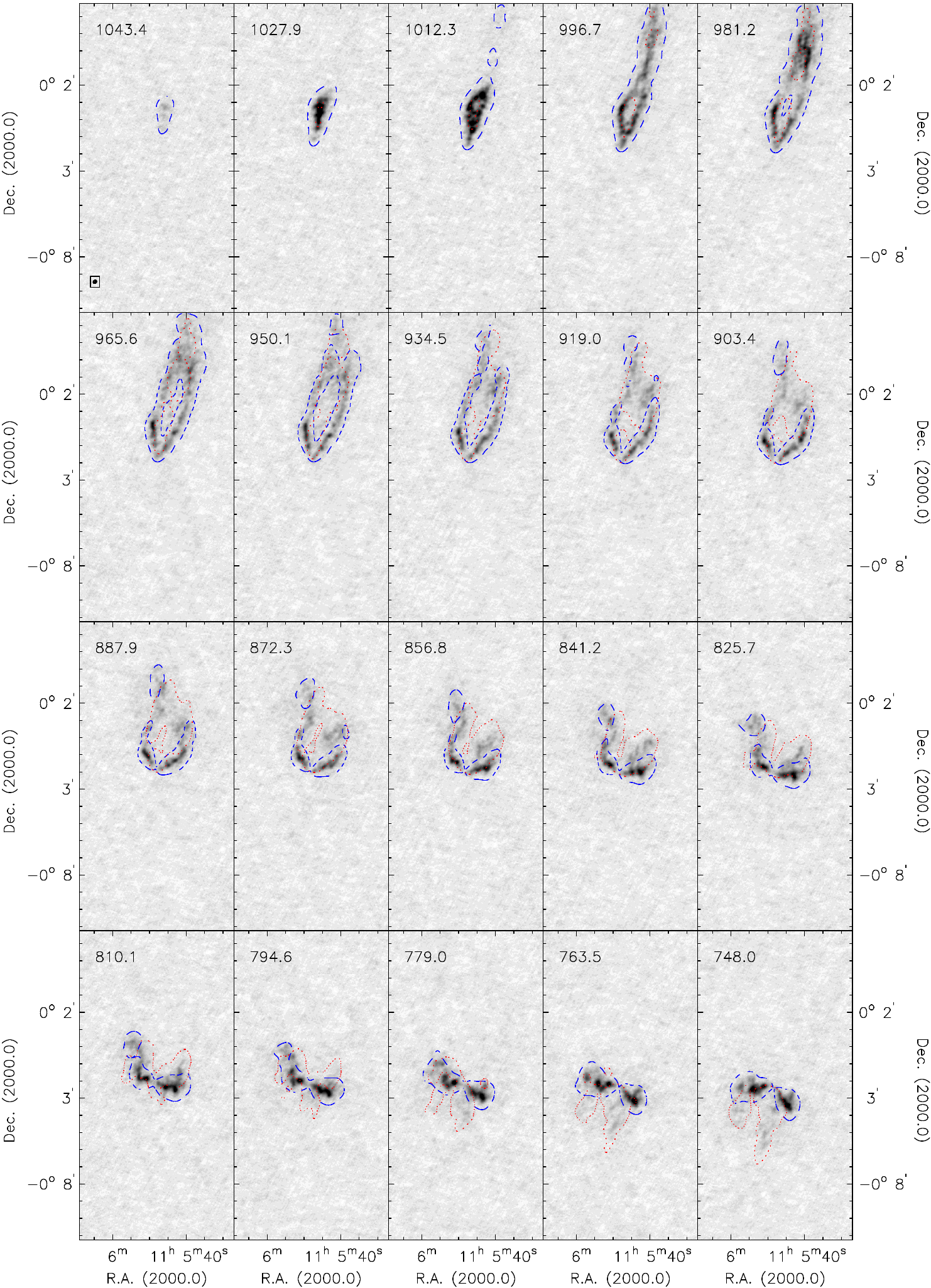}
\caption{Channel maps of the NGC~3521 \hi\ data cube from \citet{THINGS_walter}.  The greyscale range is from $2\sigma$  to $30\sigma$ where $\sigma=0.39$~mJy~beam$^{-1}$ is the r.m.s. of the noise in a line-free channel.  Blue-dashed and red-dotted contours roughly delimit the spatial distribution of the regular and anomalous \hi\ emission in each channel, and represent respective \hi\ column density levels of $\sim 9\times 10^{18}$ and $3\times 10^{18}$~cm$^{-2}$.   The heliocentric radial velocity of each channel is shown units of km s$^{-1}$ in the upper left corner.  The half-power-beam-width ($14.14$~arcsec~$\times$~11.15~arcsec) is shown in the top left panel.  Only every third channel of the data cube is shown.}
\label{channel_maps1}
\end{figure*}
\addtocounter{figure}{-1}

\begin{figure*}
\centering
\includegraphics[width=1.8\columnwidth, angle=0]{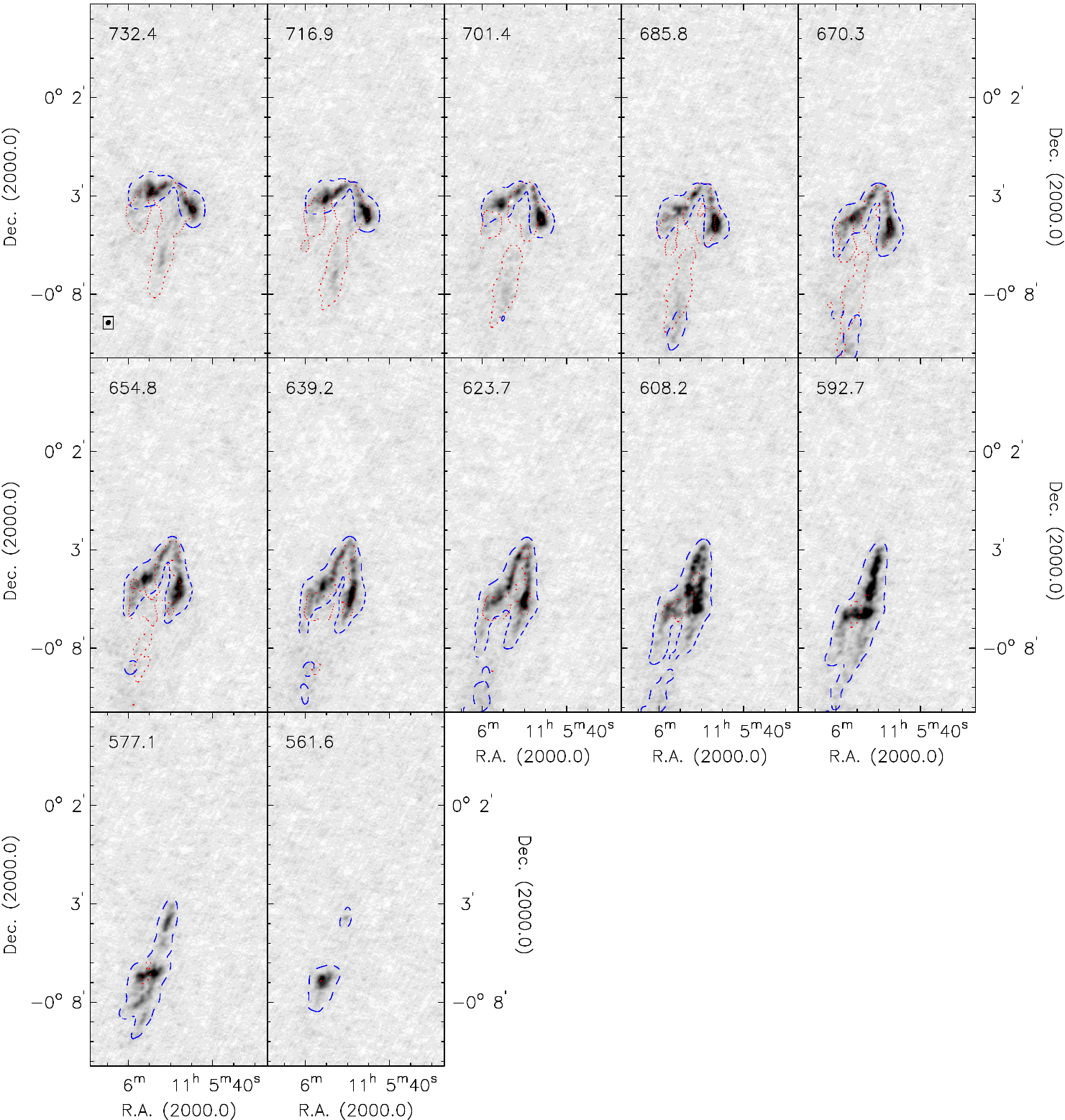}
\caption{Continued.}
\end{figure*}  

\subsection{Global profiles}\label{sec_global_profiles}
\begin{figure}
\centering
\includegraphics[width=1.\columnwidth, angle=0]{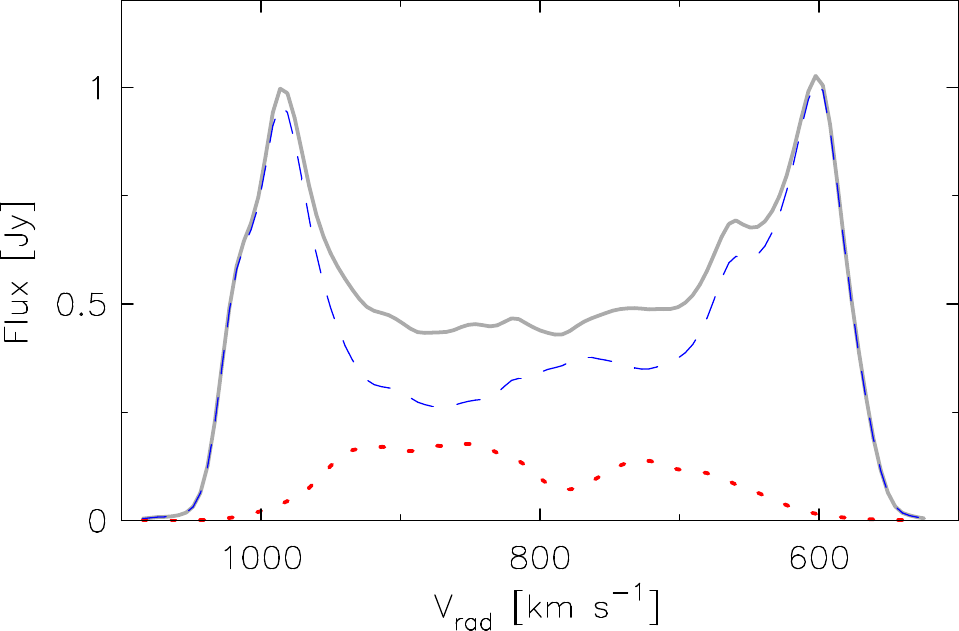}
\caption{Global profiles extracted from the decomposed \hi\ data cubes of NGC~3521.  The blue long-dashed and red short-dashed curves represent the total flux within each channel of the regular and anomalous \hi\ cubes, respectively.  The solid black curve represents the global profile of the full THINGS data cube (from which the regular and anomalous \hi\ data cubes were extracted). }
\label{global_profiles}
\end{figure}

The flux within each channel of the regular and anomalous \hi\ data cubes was summed to produce a global profile for each \hi\ mass component, shown in Fig.~\ref{global_profiles}.  Because separate regular \emph{and} anomalous \hi\ components were not identified at every position within the galaxy, the combined flux from the regular and anomalous cubes is expected to underestimate the total galaxy flux contained in the original THINGS data cube.  To correct for this the global profiles were scaled to yield a combined total \hi\ flux $\mathrm{S_{HI}} = 297.2$~Jy~\kms\ as given by \citet{THINGS_walter} for the full THINGS cube.  Adopting the corresponding total \hi\ mass of $8.02\times 10^9$~\msun\ (for a distance of 10.7~Mpc), the masses of the regular and anomalous \hi\ components in NGC~3521 are $\mathrm{M_{HI}^{reg}}=6.4\times 10^9$~\msun\ and $\mathrm{M_{HI}^{anom}}=1.5\times 10^9$~\msun, respectively.  Thus, the anomalous \hi\ constitutes 20~per~cent of the total \hi\ mass.  %For NGC~3198/2403, \citet{gentile_N3198}/\citet{ngc2403_beard} find $\sim 15/10$~per~cent of the total \hi\ mass to be contained within an anomalous gas component.

\subsection{\hi\ total intensity}
\begin{figure*}
\begin{sideways}
\begin{minipage}{23.5cm}
\includegraphics[width=0.5\columnwidth, angle=270]{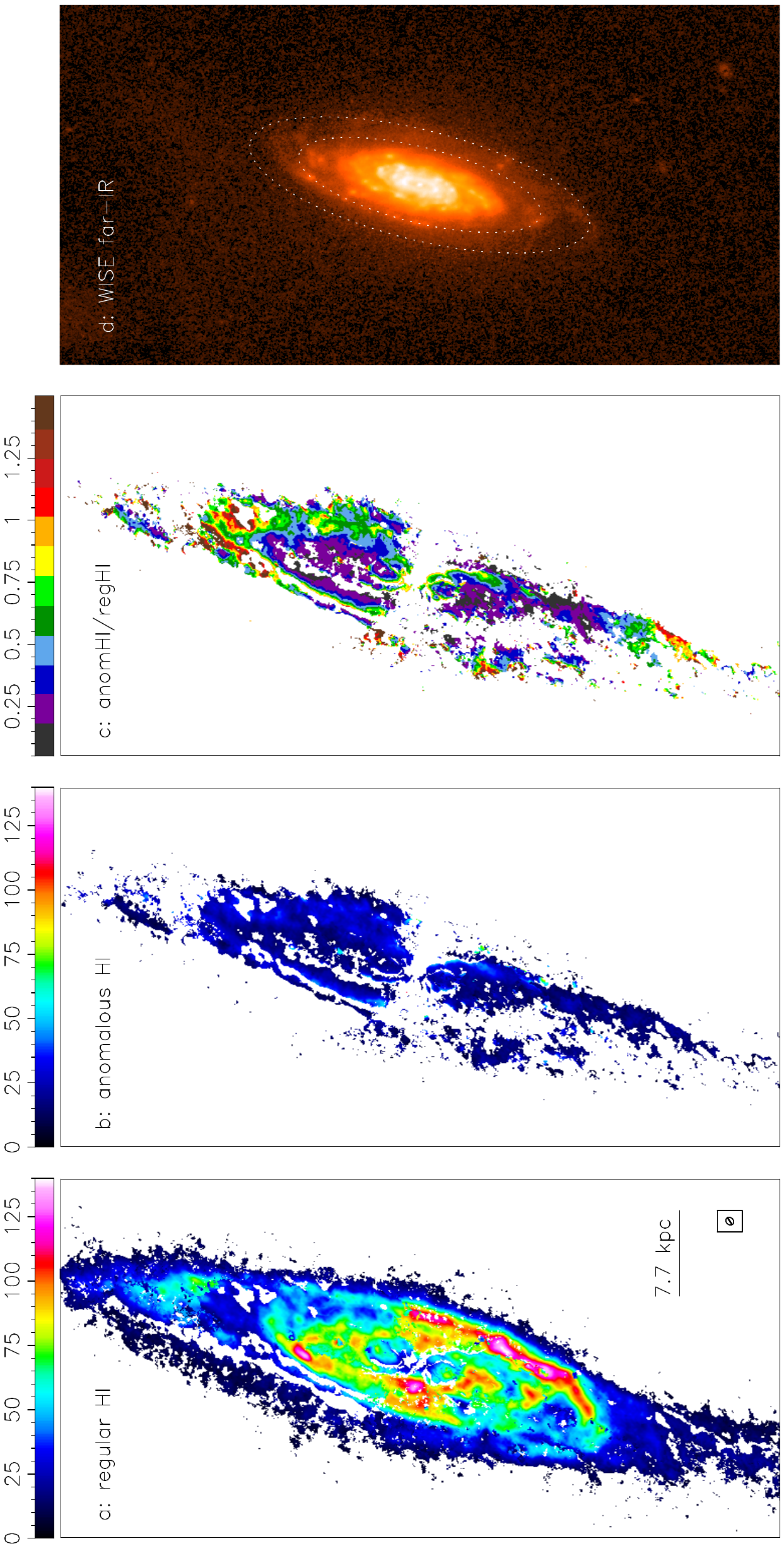}
\caption{NGC~3521 \hi\ total intensity maps. Panel \textit{a/b}: \hi\ total intensity map of the regular/anomalous \hi\ component of the galaxy.  Panel \textit{c}: Ratio of the anomalous \hi\ to the regular \hi.  Panels \textit{a} and \textit{b} use the same colour scale to represent the \hi\ column density, given in units of $10^{20}$~cm$^{-2}$ by the colour bar.  The hatched circle in the lower right corner of panel \textit{a} represents the half power beam width of the synthesised beam.  The scale bar corresponds to angular and physical lengths of 150~arcsec and 7.7~kpc, respectively.  Panel \textit{d}: WISE mid-infrared {\color{black}12}~\micron\ map.  {\color{black}The inner dotted ellipse marks the radius containing 90~per~cent of the total flux; $R_{90}=189$~arcsec.  The outer dotted ellipses marks the edge of the stellar disc at $R= 275$~arcsec.}   All images are shown on the same scale.}
\label{mom0_map}
\end{minipage}
\end{sideways}
\centering
\end{figure*}

\begin{figure}
\begin{centering}
\includegraphics[width=1.\columnwidth, angle=0]{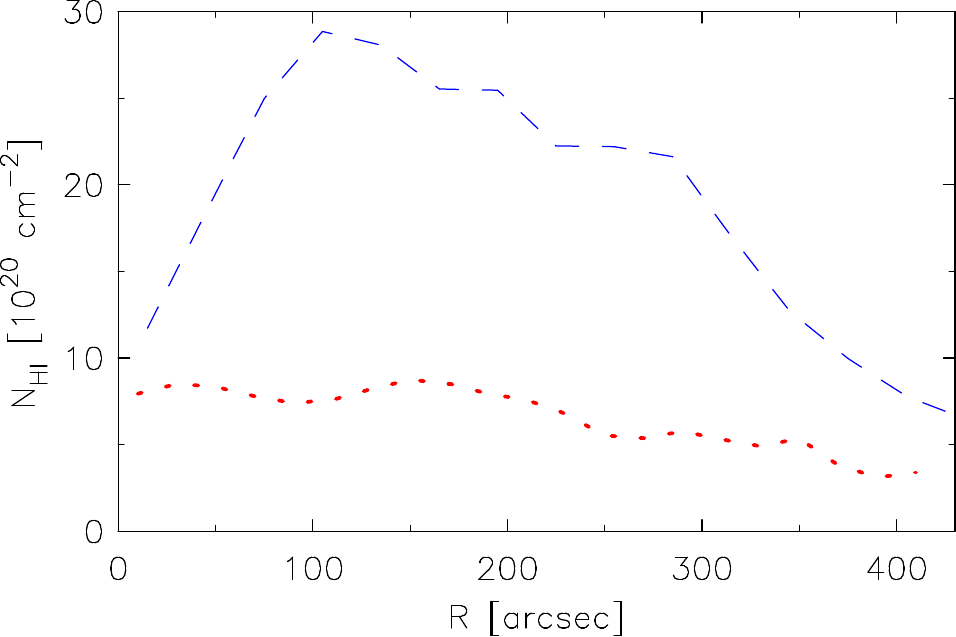}
\caption{Azimuthaly-averaged \hi\ column densities for the regular (blue long-dashed) and anomalous (red short-dashed) \hi\ components in NGC~3521.}
\label{NHI_profiles}
\end{centering}
\end{figure}

The \hi\ total intensity maps for the regular and anomalous \hi\ cubes are shown in Figs.~\ref{mom0_map}a,~b.  In all maps the blank pixels mark the positions of \hi\ line profiles for which regular and/or anomalous \hi\ components could not be unambiguously identified according to the set of criteria given in Section~\ref{sec_profile_parameterisations}.  The blank regions are therefore not necessarily indicative of a lack of a particular \hi\ component, but rather a lack of clear evidence therefor.  Azimuthally-averaged \hi\ column densities for the regular and anomalous \hi\ mass components are shown as a function of radius in Fig.~\ref{NHI_profiles}.  

The ratio of the two maps \hi\ total intensity maps (Fig.~\ref{mom0_map}c) shows the mass in anomalous \hi\ to typically be less than half of the mass in regular \hi.  However, there are regions in the galaxy where the mass in anomalous \hi\ is as much as 75~-~100~per~cent of the mass in regular \hi.

For comparison to the \hi\ maps, Fig.~\ref{mom0_map}d shows a Wide-field Infrared Survey Explorer (WISE, \citealt{WISE}) {\color{black}12}~\micron\ image of the galaxy taken from the WISE Enhanced Resolution Galaxy Atlas \citep{Jarrett_WERGA}.  The bulk of the regular \hi\ mass in NGC~3521 is clearly coincident with the stellar disc.

\subsection{Velocity fields}\label{sec_vel_fields}
Traditional intensity-weighted-mean \hi\ velocity fields approximate the line-of-sight velocities as the 1st-order moments of the line profiles.  For NGC~3521 with its asymmetric \hi\ line profiles, such a velocity parameterisation would not reliably represent the complex kinematics of the galaxy.  Because the velocity fields presented here were derived using robust parameterisations of the line profiles, they are expected to represent the true \hi\ kinematics of the galaxy much more accurately and reliably.

\begin{figure*}
\begin{centering}
\includegraphics[width=2.\columnwidth, angle=0]{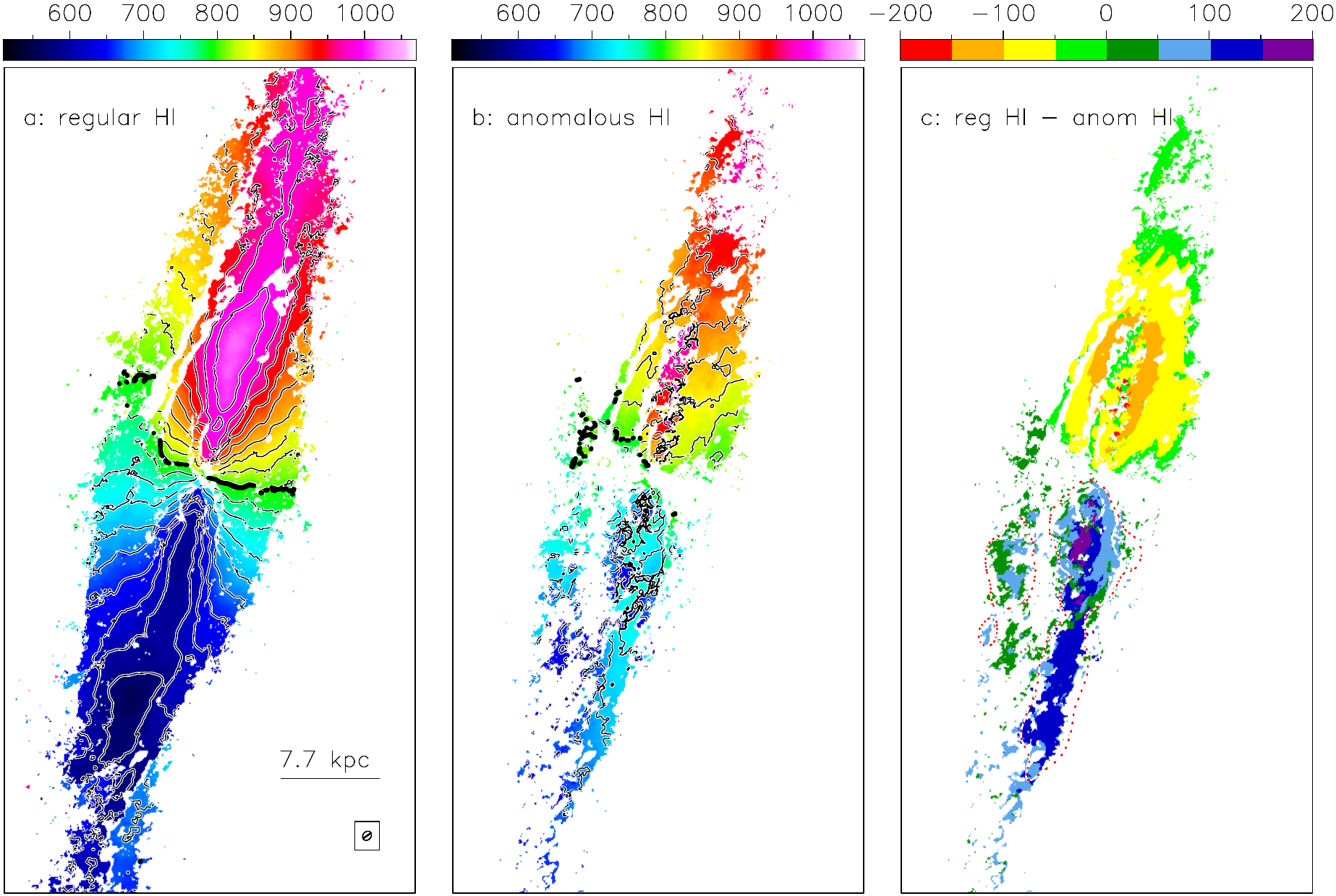}
\caption{NGC~3521 \hi\ velocity fields.  Panel~\textit{a/b}: Velocity field of the regular/anomalous \hi\ component of the galaxy.  In each panel the iso-velocity contours are separated by 25~\kms.  The thick iso-velocity contour at 798.2~\kms\ marks the systemic velocity of the galaxy.  Both panels use the same colour scale, given in units of \kms\ by the colour bar.  The hatched circle in the lower right corner of panel \textit{a} represents the half power beam width of the synthesised beam.  The scale bar corresponds to angular and physical lengths of 150~arcsec and 7.7~kpc, respectively.  Panel~\textit{c}: residual velocities obtained by subtracting the anomalous \hi\ velocity field from the regular \hi\ velocity field.  Units are given in \kms\ by the colour bar.  {\color{black}The red-dotted contours overlaid on the southern half of the galaxy are the same as those shown in the channel map at $V_{hel}=716.9$~\kms.}}
\label{mom1_map}
\end{centering}
\end{figure*}

Figures~\ref{mom1_map}a and b show the velocity fields extracted from the regular and anomalous \hi\ data cubes, respectively.  The former velocity field is clearly representative of a differentially rotating disc.  The high degree of symmetry in the iso-velocity contours about the major axis of the galaxy suggests the disc to be dominated by regular circular rotation.  The lack of any ``S-shaped'' contours provides strong evidence against the presence of a strong line-of-sight warp in the \hi\ disc.  The closed contours are indicative of a slightly declining outer rotation curve.

Despite having a low filling factor, the velocity field for anomalous \hi\ data cube is also representative of a differentially rotating disc.  With exactly the same colour scale being used for Figs.~\ref{mom1_map}a and b, it is clear that this velocity field represents a slow-rotating \hi\ component.  

{\color{black}For each \hi\ line profile, the deviation velocity $V_{dev}$ can be defined as the difference between the parameterised line-of-sight velocities of the regular and anomalous \hi\ components.  The deviation velocity field} generated by subtracting the anomalous \hi\ velocity field \textit{from} the regular \hi\ velocity field is shown in Fig.~\ref{mom1_map}c.  {\color{black}In the southern portion of the map there is a large stripe of coherent residuals in the range 50~\kms~$\lesssim V_{dev}\lesssim$~100~\kms\ extending from the centre of the galaxy all the way to its edge.  This feature spatially coincides with the anomalous \hi\ emission seen in the channel maps at a heliocentric velocity $V_{hel}\sim716.9$~\kms\ (represented by red contours in Fig.~\ref{mom1_map}c).  It \emph{does not} coincide with the extended stripe of high-intensity emission seen in the south-western portion of the regular \hi\ total intensity map (Fig.~\ref{mom0_map}a).  In the northern half of the galaxy there is a ring of coherent residuals in the range $-150$~\kms~$\lesssim V_{dev}\lesssim -100$~\kms.  Both this feature and the stripe in the south  seem to occur at locations in the galaxy where the anomalous \hi\ constitutes less than $\sim20$~per~cent of the total flux in an \hi\ line profile.}

\begin{figure}
\begin{centering}
\includegraphics[width=1.\columnwidth, angle=0]{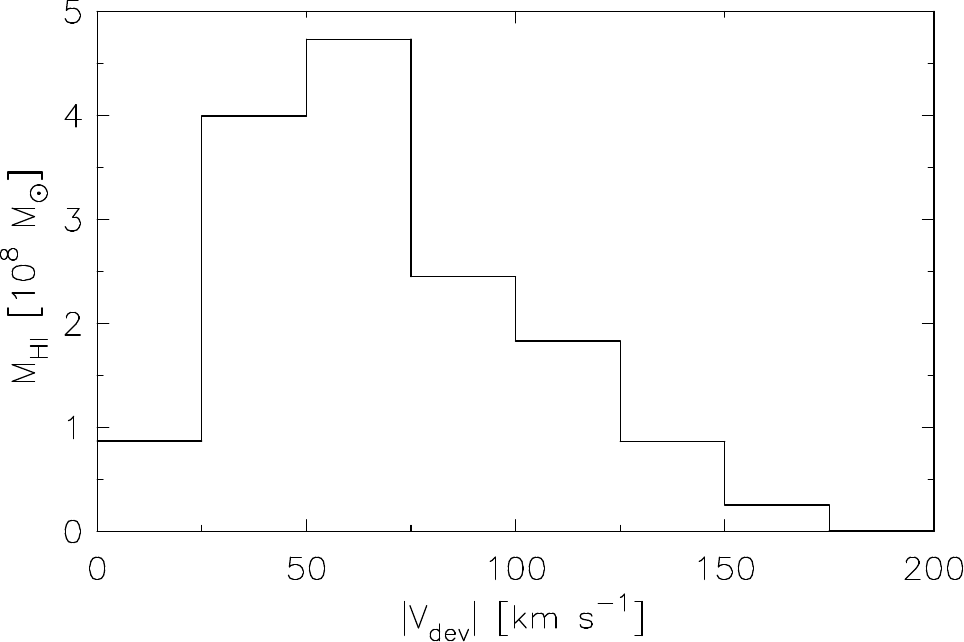}
\caption{{\color{black}Distribution of anomalous \hi\ mass as a function of absolute deviation velocity.  Approximately 80~per~cent the anomalous \hi\ mass in NGC~3521 has a rotation speed that is less than 100~\kms\ deviant from the rotation speed of the thin \hi\ disc.}}
\label{Vdevfig}
\end{centering}
\end{figure}

{\color{black}Figure~\ref{Vdevfig} shows the distribution of anomalous \hi\ mass as a function of absolute deviation velocity $|V_{dev}|$.  Approximately 80~per~cent ($1.2\times 10^9$~\msun) of the anomalous \hi\ mass in NGC~3521 has a rotation speed that is less than 100~\kms\ deviant from the rotation speed of the regular \hi\ in the thin disc.  Roughly 60~per~cent of the anomalous \hi\ is 25~-~75~\kms\ deviant.}

%TR MODELS----------------------------------------------------------------------------------------------------------------------------------------------------------------------------------------------
\section{Dynamical modelling}\label{sec_modelling}
This section aims to describe and present two sets of dynamical models derived for each of the regular and anomalous \hi\ components in NGC~3521.  Tilted ring models were fitted to the velocity fields, yielding rotation curves.  The rotation curves were subsequently used to generate full three-dimensional models for each of the regular and anomalous \hi\ data cubes in order to quantify the distribution of the \hi\ mass.  

\subsection{Tilted ring modelling}\label{TRmodeling}
One of the standard methods of producing a rotation curve for a galaxy disc involves modelling the velocity field as a set of concentric rings within which the gas moves on circular orbits - the so-called tilted ring model \citep{Rogstad1974}.  The clear signatures of circular rotation seen in the \hi\ velocity fields of NGC~3521 make them well suited to tilted-ring modelling.

Most generally, each tilted ring is defined by a set of parameters: central coordinates $X_c$ and $Y_c$, inclination $i$, position angle $PA$ (in the sky plane\symbolfootnote[3]{Measured anti-clockwise from north to the receding semi-major axis.}), systemic velocity $V_{sys}$ and $V_{exp}$ and $V_{rot}$ which are the respective expansion and rotation speeds of the material moving within the ring.  A least-squares fit of the following function is made to the velocity field:
\begin{equation}
V_{los}(x,y)=V_{sys}+\sin i(V_{rot}\cos\theta+V_{exp}\sin\theta),
\label{vlos}
\end{equation}
where $V_{los}$ is the heliocentric line-of-sight velocity, $x$ and $y$ are rectangular coordinates on the sky and $\theta$ is the angle from the major axis in the galaxy plane.  $\theta$ is related to the position angle $PA$ in the sky plane by
\begin{eqnarray}
\cos(\theta)&=&{-(x-X_c)\sin PA+(y-Y_c)\cos PA\over R},\\
\sin(\theta)&=&{-(x-X_c)\cos PA+(y-Y_c)\sin PA\over R\cos i},
\end{eqnarray}
where $R$ is the radius of the ring in the galaxy plane.

\subsubsection{Method}
The Groningen Image Processing SYstem ({\sc gipsy}, \citealt{gipsy}) task {\sc rotcur} \citep{begemann_phd_thesis} was used to fit tilted ring models to the regular and anomalous \hi\ velocity fields.  Rings of width $\Delta R=15$~arcsec were used to ensure that adjacent rings were largely independent of one another.  When fitting Eqn.~\ref{vlos} to the data, each datum was weighted by $|\cos\theta|$ so that points closer to the major axes hold more weight.  All points within 10$^{\circ}$ of either side of the minor axes were excluded from the fit.  Both sides of the galaxy were used for the fitting.  The tilted ring models were fitted in an iterative fashion.  Given that  the photometric and dynamical centres of NGC~3521 are expected to lie close to one another (\citealt{trachternach_THINGS}), the centre positions of the rings were always fixed to the dynamical centre position of the galaxy (given in Section~\ref{sec_intro}).  In all runs $V_{sys}=798.2$~\kms\ was fixed.  

\subsubsection{Results}
\begin{figure*}
\centering
\includegraphics[width=2.\columnwidth, angle=0]{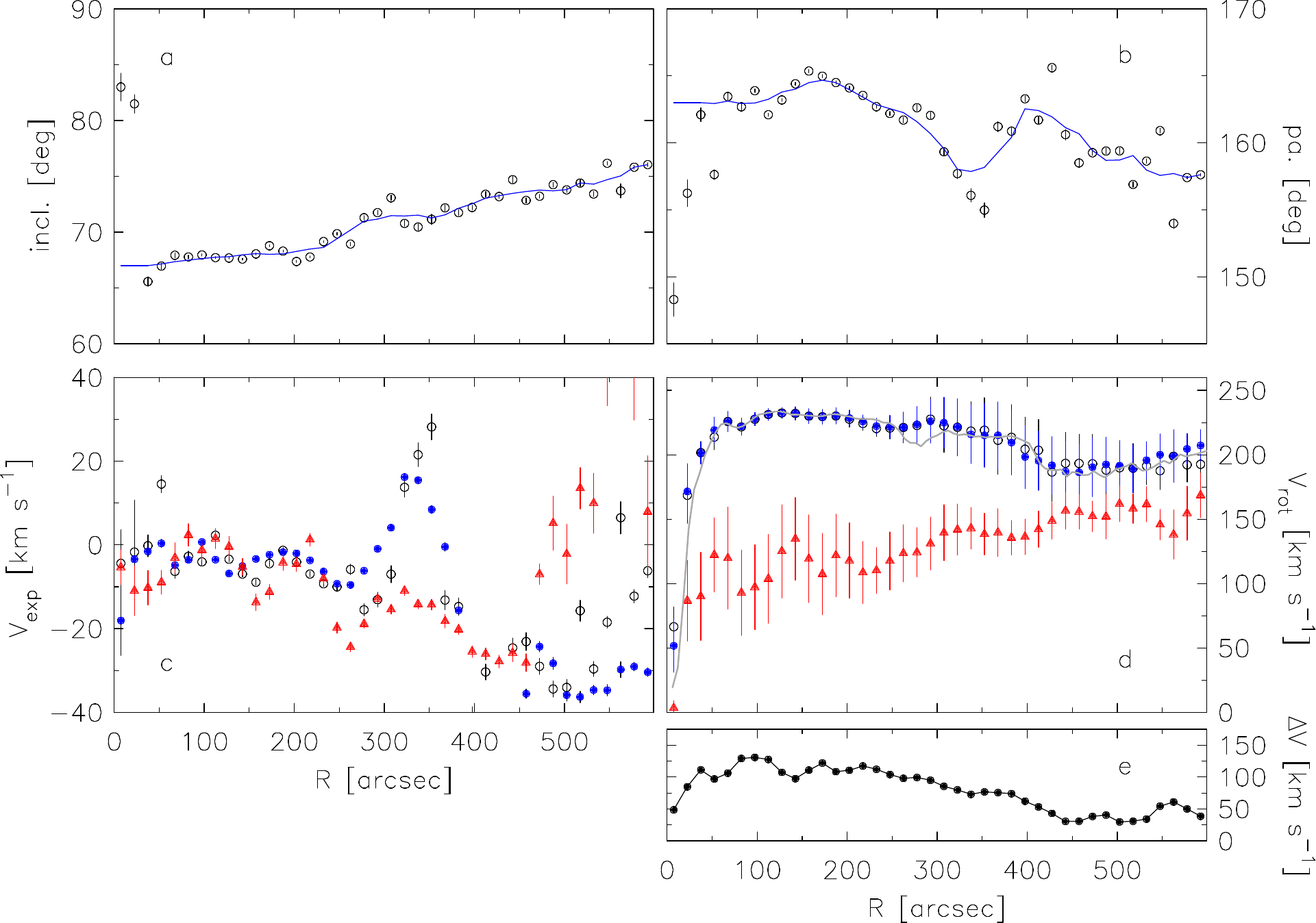}
\caption{Tilted ring models for the regular and anomalous \hi\ velocity fields.  Panels $a$, $b$, $c$ and $d$ show the radial runs of disc inclination, disc position angle, expansion (radial) velocity of the \hi\ and rotation (circular) velocity of the \hi, respectively.  Open black circles in panels $a$~-~$d$ show the radial runs for the first (unconstrained) fit to the regular \hi\ velocity field.  {\color{black}The solid blue curves in panels $a$ and $b$ show the boxcar smoothed versions of the profiles used to derive the final tilted ring models for the regular and anomalous \hi\ velocity fields.  The final runs of $V_{exp}$ and $V_{rot}$ for the regular and anomalous \hi\ velocity fields are shown as red-filled triangles and blue-filled circles, respectively.}  The solid grey curve in panel $d$ shows the rotation curve derived by \citet{THINGS_deblok}.  The difference between the regular and anomalous \hi\ rotation curves is shown in panel $e$.}
\label{TR_results}
\end{figure*}

The first iteration carried out for the regular \hi\ velocity field had $i$, $PA$, $V_{exp}$ and $V_{rot}$ freely varying with radius.  The resulting radial profiles are shown as black open circles in Figs.~\ref{TR_results}a~-~d.  The radial run of inclinations suggests the regular \hi\ disc to be very mildly warped.  The variation in position angle is generally less than $\sim10$~degrees.  The $i$ and $PA$ profiles were boxcar smoothed (solid blue lines in Figs.~\ref{TR_results}a,~b) and fixed in order to derive the final radial runs of $V_{exp}$ and $V_{rot}$ (filled-blue circles in Figs.~\ref{TR_results}c,~d).  The rotation curve derived in this work is very similar to the one generated by \citet{THINGS_deblok} (solid grey curve in Fig.~\ref{TR_results}d).  %The only significant differences occur at radii $R\lesssim 50$~arcsec where the newly generated curve lies above the \citet{THINGS_deblok} curve.

The radial runs of $PA$ and $i$ generated for the regular \hi\ velocity field (blue-filled curves in Figs.~\ref{TR_results}a,~b) were applied  to the anomalous \hi\ velocity field in order to generate $V_{exp}$ and $V_{rot}$ radial profiles (red-filled triangles in Figs.~\ref{TR_results}c,~d).  In terms of circular rotation speed, the models show the anomalous \hi\ to lag the regular \hi\ at all radii.  The largest lag of $\sim120$~\kms\ occurs near the knee of the rotation curve ($R\sim 100$~arcsec, Fig.~\ref{TR_results}e).  Beyond the knee the two rotation curves begin to converge but are still always separated by $\sim 25$~-~100~\kms.  The anomalous \hi\ in NGC~3521 is  clearly a slow-rotating component - lagging the rotation of the thin \hi\ disc. 

{\color{black}For radii $R\lesssim 275$~arcsec the expansion velocities for both tilted ring models are roughly similar, and are suggestive of radial inflows at the level of 0~-~10~\kms\ (cf.~\citealt{wong}).  The expansion velocities for the two models largely diverge for radii $R\gtrsim 275$~arcsec.  This radius marks the approximate edge of the stellar disc, and is shown as an ellipse in Fig.~\ref{mom0_map}d.}  %\citet{briggs_warps_1990} suggests \hi\ discs of galaxies to be warped beyond the edge of the stellar disc.  

\subsection{Residuals}
\begin{figure}
\centering
\includegraphics[width=1.\columnwidth, angle=0]{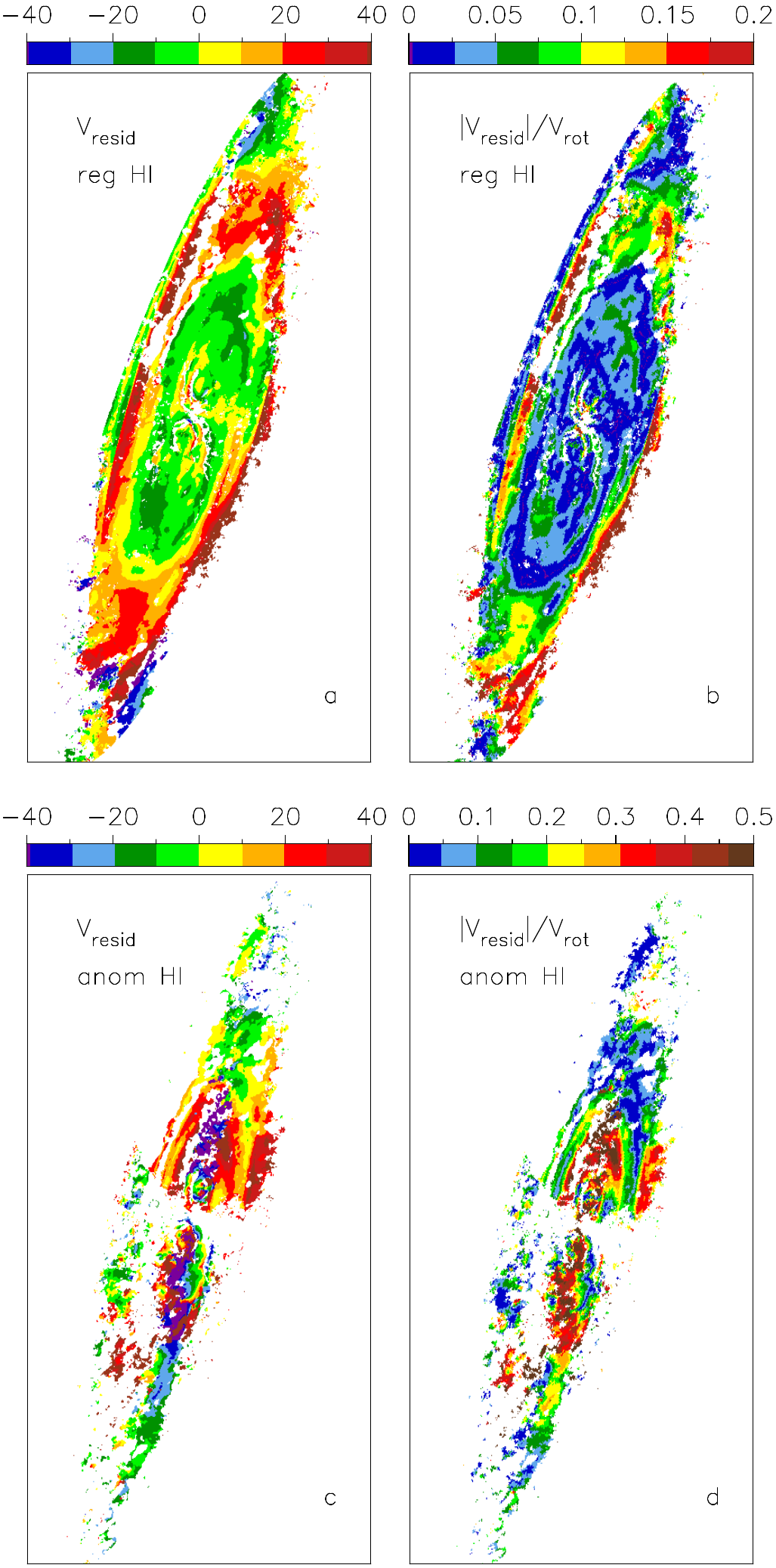}
\caption{Panels $a,c:$ Residual velocity fields, generated by subtracting the tilted ring models from the data.  Units are given in \kms\ by the colour bars.  Panels $b,d:$ Ratio of the absolute residual velocities to the corresponding rotation velocities.}
\label{resids}
\end{figure}

{\color{black}The tilted ring models were converted into velocity fields which were subtracted from the data to produce residual velocity fields.  These fields for the regular and anomalous \hi\ velocity fields are shown in Fig.~\ref{resids}a and Fig.~\ref{resids}c, respectively.  The majority of the regular \hi\ residual velocities lie in the range $-10$~\kms~$\lesssim V_{resid}\lesssim 10$~\kms, suggesting the tilted ring model to be a good fit to the data.  The residuals in the outer disc are noticeably larger; up to $|V_{resid}|\gtrsim 20$~\kms\ in magnitude.  However, compared to the typical rotation speed of the outer disc, the residuals are relatively small.  Figure~\ref{resids}b is a map, for the regular \hi\ velocity field, showing the ratio of the absolute  residual velocities to the corresponding circular velocities ($V_{rot}$, according to the titled ring model).  Even the largest absolute residual velocities are less than $\sim 20$~per~cent of the circular rotation speed.  In fact, over the vast majority of the disc the absolute residuals are less than $\sim 5$~-~10~per~cent of the circular rotation speed.

The situation is slightly different for the anomalous \hi\ tilted ring model.  Figure~\ref{resids}c shows it to have the largest residual velocities ($|V_{resid}|\sim$~20~-~30~\kms) occurring at inner radii, with smaller residuals ($|V_{resid}|\lesssim 10$~\kms) in the outer disc.  Figure~\ref{resids}d shows the residual velocities in the outer disc to be a factor of $\sim 10$ smaller than the rotation speed of the anomalous \hi. At inner radii, the residual velocities are $\sim 20$~-~40~per~cent the magnitude of the rotation speed.}

\subsection{Three-dimensional modelling}
While the tilted ring models strongly suggest the anomalous \hi\ in NGC~3521 to be a slow-rotating component, they provide no insight into its possible distribution.  Three-dimensional models of the \hi\ data cubes can be used to check if the anomalous \hi\ is contained in a thin disc or whether its more likely distributed in a thick disc or halo.  Furthermore, they allow the results of the tilted ring models to be directly compared to the \hi\ data cubes, rather than simply the \hi\ velocity fields.  This section is dedicated to producing separate three-dimensional models of the regular and anomalous \hi\ data cubes, combining them and comparing the resulting cube to the observations.  

\subsubsection{Method}
The {\sc gipsy} task {\sc galmod} was used to generate three-dimensional model cubes.  Assuming axisymmetry, the routine distributes `\hi\ clouds' into a set of tilted rings.  The orientation parameters of each ring must be provided as well as the rotation velocity, velocity dispersion and column density of the \hi\ within the ring.  For each \hi\ cloud placed in a ring, a velocity profile around the position of the cloud is built using the systemic and rotation velocities of the tilted ring.  

Separate models were produced for each of the regular and anomalous \hi\ data cubes.  The \hi\ surface density profiles shown in Fig.~\ref{NHI_profiles} were used together with the boxcar smoothed $PA$ and $i$ profiles shown in Figs.~\ref{TR_results}a and b (solid blue curves).  Only circular velocities were modelled using the rotation curves derived in Section~\ref{TRmodeling} (Fig.~\ref{TR_results}d).  An \hi\ velocity dispersion of 10~\kms\ was used for all models together with a systemic velocity $V_{sys}=798.2$~\kms.  As was done for the tilted ring models, the centre positions of the rings were fixed to the dynamical centre position of the galaxy.  A Gaussian profile was used for the \hi\ density distribution in the direction perpendicular to the plane of the rings.  For the model of the regular \hi\ a constant scale-height of $\sim 260$~pc was used.  No flaring of the \hi\ disc at outer radii was incorporated.

\subsubsection{Results}
\begin{figure*}
\begin{sideways}
\begin{minipage}{22.5cm}
\includegraphics[width=0.48\columnwidth, angle=270]{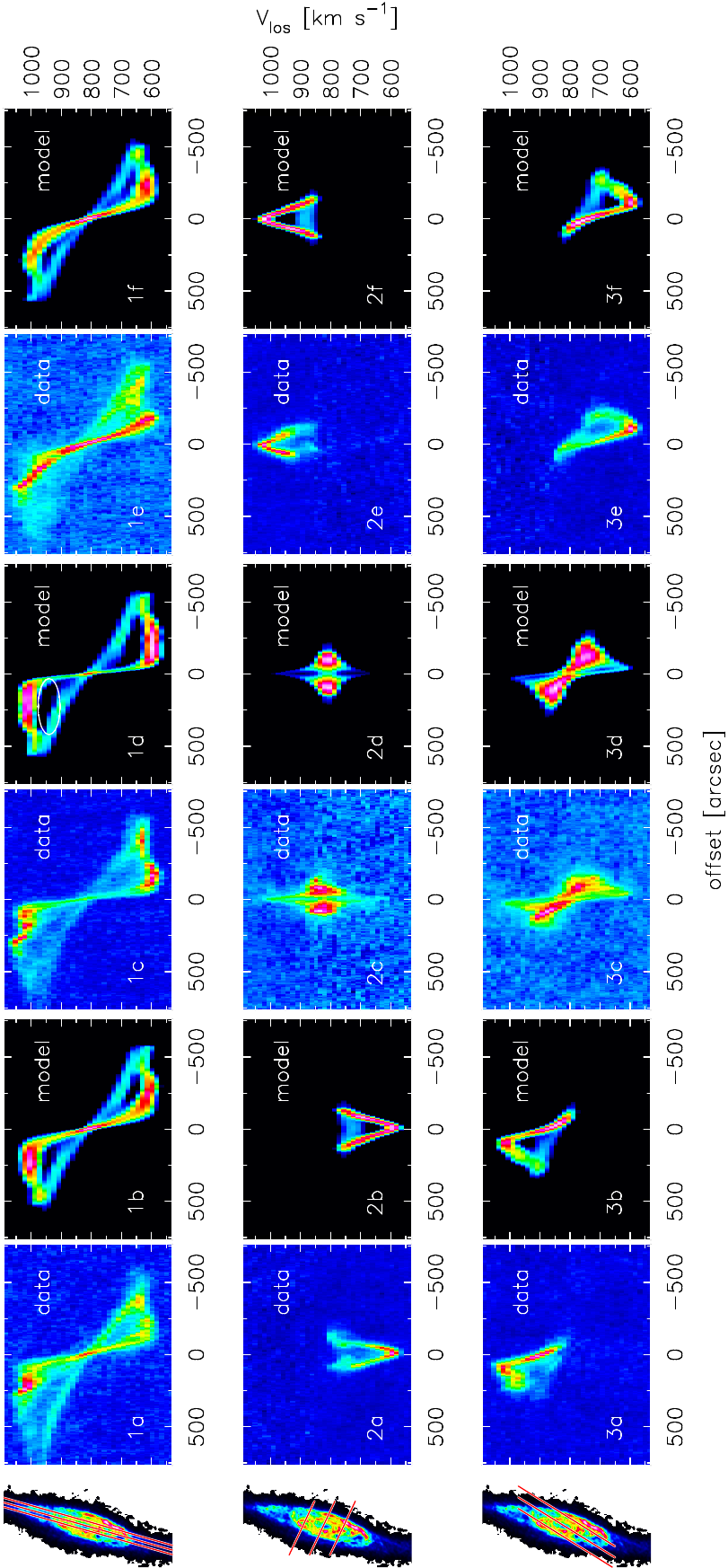}
\caption{Position-velocity slices extracted from the \hi\ data cube and the model cube.  Each of rows 1~-~3 show three different slices extracted parallel, perpendicular and at 45~degrees to the major axis of the \hi\ disc, respectively.  The slices are shown as red lines in the \hi\ total intensity maps.  Each slice is presented as a data-model pair.  The \hi\ total intensity map is taken from the THINGS pubic data repository (http://www.mpia-hd.mpg.de/THINGS/Data.html).  The white ellipse shown in panel $1d$ roughly delimits the velocity gap between the regular and anomalous \hi\ components of the model cube.  For the model cube presented here, the anomalous \hi\ is treated as being distributed in a thick ($\sim 3.5$~kpc), slow-rotating disc with no vertical gradient in the direction perpendicular to the \hi\ disc.  No thin-disc models for the anomalous \hi\ can match the observations.}
\label{slice_panel}
\end{minipage}
\end{sideways}
\centering
\end{figure*}

Having generated separate models for the regular and anomalous \hi\ data cubes, they were added together to produce a final combined model cube to compare to the observations.  Figure~\ref{slice_panel} shows the modelling results in the form of position-velocity slices extracted parallel, perpendicular and at 45~degrees to the major axis of the \hi\ disc (rows 1, 2 and 3).  A data-model pair is presented for each slice.

The main characteristics of the regular \hi\ are easily reproduced by a simple thin-disc model.  However, no thin-disc models can reproduce the observed spatial and spectral distribution of the anomalous \hi.  Only models incorporating an \hi\ scale-height of $\sim~3$~-~4~kpc can match the characteristics of the data.  A scale-height of 3.5~kpc for the anomalous \hi\ was used to generate the model shown in Fig.~\ref{slice_panel}.  The model cubes are axisymmetric by construction, and therefore cannot match the kinematic asymmetry of the galaxy.  Nevertheless, they successfully reproduce the main features of the observations.  In each of the position-velocity slices shown in Fig.~\ref{slice_panel}, the anomalous \hi\ in the model cubes shows up as diffuse emission that is noticeably separated from the high-concentration emission from the \hi\ disc.   

The models presented in this work treat the anomalous \hi\ as being smoothly distributed in a thick disc with no vertical gradient in its rotation speed.  This simplifying assumption yields a velocity gap between the regular and anomalous \hi\ components, as clearly seen in position-velocity slices extracted parallel to the major axis of the galaxy.  In Fig.~\ref{slice_panel}-1d the gap is roughly delimited by a white ellipse.  In reality there is expected to be a vertical velocity gradient in the rotation speed of the halo - the inner halo rotates faster than the outer halo (e.g.~\citealt{oosterloo_NGC891}), leading to a filling-in of the velocity gap seen in the position-velocity slices extracted from the models.  The lack of a velocity gap in panels 1a, c and e in Fig.~\ref{slice_panel} provides observational evidence for a vertical gradient in the rotation speed of the halo of NGC~3521.

The fact that the simple models presented in this work do a good job of reproducing the main features of the observations  demonstrates two things about the anomalous \hi: 
\begin{enumerate}
\item It is not co-rotating with the thin \hi\ disc
\item It is not distributed in a thin \hi\ disc; it has to be extra-planar.
\end{enumerate}
These findings alone paint a picture of a galaxy with the bulk of its \hi\ mass residing in a thin \hi\ disc, yet with a sizeable fraction of it distributed in a diffuse, slow-rotating halo.

\section{Origin of anomalous \hi}\label{anomHI_origin}
There are two main, generally accepted, mechanisms by which gas can enter the halo of a galaxy.  
\subsection{Galactic fountain}
Galactic fountains \citep{galactic_fountain} play a major role in building up a galaxy's halo.  In this model, star formation ejects gas from the disc of the galaxy.  The material moves into the halo, cools and subsequently rains back down onto the disc.  Edge-on galaxies for which the extra-planar \hi\ can be directly observed (e.g. NGC~891) often show it to be concentrated close to the stellar disc.  \citet{fraternali_binney_2006} generate models capable of reproducing the observed vertical distribution of extra-planar gas in NGC~891 for an energy input corresponding to a small fraction of the energy released by supernovae.  Given that NGC~3521 and NGC~891 have comparable star formation rates of 3.3~\msun~yr$^{-1}$  and 3.8~\msun~yr$^{-1}$  \citep{popescu_2004} as well comparable sizes, asymptotic circular velocities and halo masses, {\color{black}it may be the case} that a galactic fountain is responsible for most of the gas in the halo of NGC~3521.  

\begin{figure}
\centering
\includegraphics[width=1.\columnwidth, angle=0]{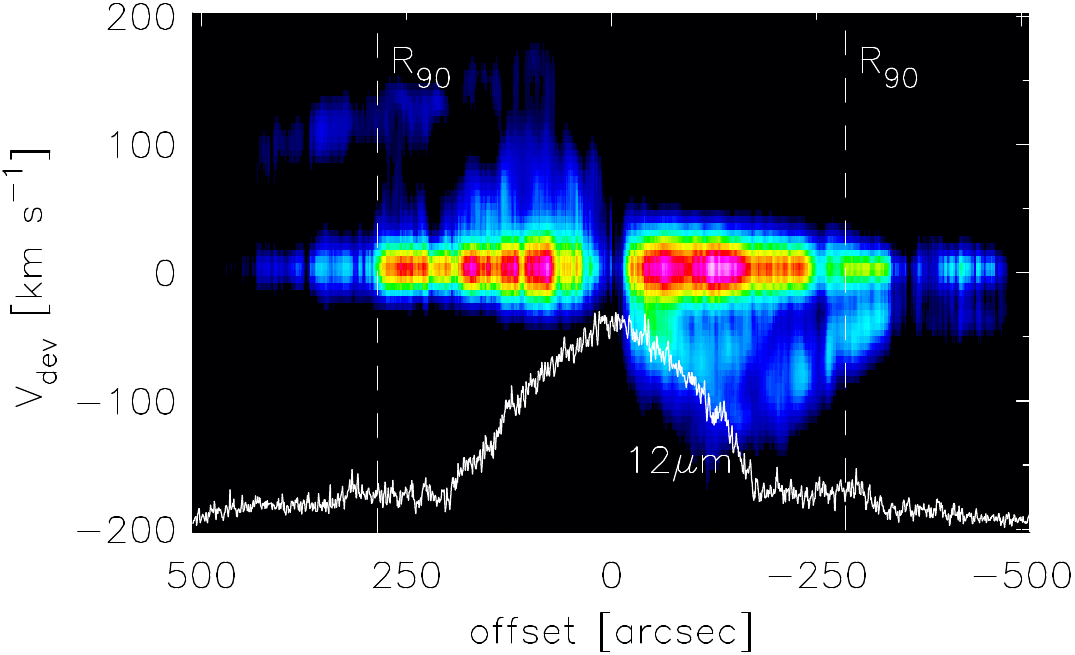}
\caption{{\color{black}Position-velocity slice extracted parallel to the major axis of the galaxy in the de-rotated \hi\ cube.  More precisely, major axis position-velocity slices were extracted at each point on the minor axis and summed together to produce the image shown here.  The broad band of de-rotated \hi\ emission at $V_{dev}=0$~\kms\ corresponds to the thin \hi\ disc in NGC~3521.  The remaining emission in the range -100~\kms~$\lesssim~V_{dev}~\lesssim$~100~\kms\ is the anomalous \hi.  The white curve represents the strip-integrated 12~\micron\ emission (in arbitrary units).  The vertical dashed lines mark the radius containing 90~per~cent of the 12~\micron\ emission; $R_{90}=189$~arcsec.  Clearly evident is the fact that the vast majority of anomalous \hi\ in NGC~3521 is spatially coincident with the inner regions of the disc where the star formation is highest.}}
\label{Vdev}
\end{figure}

{\color{black} \citet{boomsma_6946} carry out an extensive study of the high-velocity \hi\ the detect in NGC~6946.  They present a very effective method that demonstrates the anomalous \hi\ in NGC~6946 to be related to stellar feedback.  Their technique is applied here to the case of NGC~3521.  The method involves de-rotating the \hi\ data cube so that all of the regular \hi\ lies at the systemic velocity.  Each line profile in the NGC~3521 cube with parameterised regular \emph{and} anomalous \hi\ components was shifted to place the regular \hi\ component at the systemic velocity (798.2~\kms).  As \citet{boomsma_6946} explain, this removes the systematic motion (disc rotation) and results in a data cube where the third axis is no longer a heliocentric velocity axis, but rather a deviation velocity ($V_{dev}$) axis.  

Position-velocity slices aligned parallel to the major axis of the de-rotated galaxy were extracted at each position along the minor axis and summed together.  The resulting integrated position-velocity slice is shown in Fig.~\ref{Vdev}.  The broad band of de-rotated \hi\ at $V_{dev}=0$~\kms\ corresponds to the thin \hi\ disc in NGC~3521.  Most of the anomalous \hi\ lies in the range -100~\kms~$\lesssim~V_{dev}~\lesssim$~100~\kms (as already shown in Sec.~\ref{sec_vel_fields}) and occurs at galactocentric radii $R~\lesssim~250$~arcsec. } 

{\color{black}To compare the radial distribution of anomalous \hi\ in NGC~3521 to the star formation, emission in the WISE 12~\micron\ image of the galaxy (Fig.~\ref{mom0_map}d) was also strip-integrated along the minor axis.  \citet{Jarrett_WERGA} explain how the 12~\micron\ data are ``sensitive to polycyclic aromatic hydrocarbon emission arising from the photon-dominated regions located at the boundaries of \hii\ regions and molecular clouds''.  The 12~\micron\ emission is therefore closely linked to the star formation activity.  The strip-integrated 12~\micron\ emission of NGC~3521 is shown as a white curve in Fig.~\ref{Vdev}.  The vertical dashed lines in Fig.~\ref{Vdev} represent the radius containing 90~per~cent of the 12~\micron\ emission; $R_{90}=189$~arcsec.  This radius is marked with a white ellipse in Fig.~\ref{mom0_map}d.  Figure~\ref{Vdev} clearly shows the vast majority of anomalous \hi\ in NGC~3521 to be spatially coincident with the inner regions of the disc where the star formation is highest.  This finding provides strongly suggestive evidence for the anomalous \hi\ in the halo of NGC~3521 having been driven out of the disc by stellar feedback.}

\subsection{Accretion from intergalactic medium}
Although they play a main role, galactic fountains cannot account for all of the observed properties of extra-planar gas in galaxies.  Accretion of cold gas from the intergalactic medium (IGM) is also known to contribute to the gas content of galaxy halos.  The presence of halo structures with very high kinetic energy requirements ($\gtrsim 10^4$ supernovae) suggest them to originate externally \citep{sancisi_review}.  Furthermore, the material is oftentimes observed not to be dynamically linked to the disc.  Estimates for accretion rates vary from roughly 10~-~100~per~cent the star formation rate of spiral galaxies \citep{sancisi_review, fraternali_binney_2006}.  Some of the halo gas in NGC~3521 would {\color{black} have originated in the IGM.}

\section{Discussion}\label{sec_discussion}
The various models of NGC~3521 presented in this work suggest the anomalous \hi\ emission in the galaxy to be distributed in a diffuse, slow-rotating halo with a thickness of a few kpc.  Although this sort of scenario may seem extreme, the findings presented here are by no means unique to this galaxy.  Dynamical \hi\ studies of other nearby late-type galaxies have revealed similar results.  

\citet{ngc2403_beard} use VLA \hi\ observations of the nearby spiral galaxy NGC~2403 to demonstrate the presence of a faint, extended and kinematically anomalous component.  They estimate the mass of the anomalous component to be $\mathrm{M_{HI}}\sim 3\times 10^8$~\msun, about 10~per~cent of the total \hi\ mass.  They also generate a velocity field and rotation curve for the anomalous \hi, showing it to lag the circular rotation speed of the thin disc by 25~-~50~\kms.  \citet{ngc2403_beard} suggest the anomalous \hi\ to lie in a thick and clumpy \hi\ layer characterised by slower rotation and inflow motion toward the centre of the galaxy.   

In the edge-on galaxy NGC~891, deep \hi\ observations presented by \citet{oosterloo_NGC891} reveal the presence of a huge gaseous halo (M$\mathrm{_{HI}}\sim 1.2\times 10^9$~\msun) containing almost 30~per~cent of the \hi\ and extending vertically as far as 22~kpc.  The overall kinematics of the halo gas are characterised by differential rotation lagging with respect to that of the disc.  \citet{Swaters_NGC891} show the halo gas to rotate 25~-~100~\kms\ more slowly than the gas in the plane - very similar to the result for NGC~3521 presented in this work.  

{\color{black}\citet{boomsma_6946} use very deep 21~cm observations from the Westerbork Synthesis Radio Telescope to study the distribution and kinematics of the neutral gas in NGC~6946.  They detect high-velocity \hi\ (up to $\sim~100$~\kms) which they conclude to be extra-planar.  The overall kinematics of the anomalous gas is characterised by slow rotation (relative to the thin \hi\ disc).  The authors provide very compelling evidence for the amount of \hi\ mass missing from holes in the \hi\ disc to be comparable to the mass of high-velocity \hi\ in the halo, supporting their hypothesis that the two phenomena are related by means of a galactic fountain.}

{\color{black}For the moderately inclined spiral galaxy NGC~2997, \citet{hess_2009} find evidence for a galactic fountain as well as accretion of material from the IGM.  They show the galaxy to have a slow-rotating, thick HI disc consisting of 16 - 17~per~cent of the total \hi\ mass, and with a vertical velocity gradient of 18 - 31~\kms~kpc$^{-1}$.  The authors estimate a lower limit of 1.2~\msun~yr$^{-1}$ for the accretion rate of material from the IGM.}

\citet{gentile_N3198} obtain deep \hi\ observations of the spiral galaxy NGC~3198 as part of the Westerbork Accretion in LOcal GAlaxieS survey (HALOGAS, \citealt{HALOGAS}).  They present a best-fitting model of the data featuring a thick \hi\ disc with a scale-height of $\sim~3$~kpc and an \hi\ mass of about 15~per~cent of the total \hi\ mass ($\mathrm{M_{HI}}=1.08\times 10^{10}$~\msun).  They show the thick \hi\ layer to be not only slow-rotating, but to decrease in rotation speed with height at a rate of roughly 7~-15~\kms~kpc$^{-1}$.

%SUMMARY----------------------------------------------------------------------------------------------------------------------------------------------------------------------------------------------
\section{Summary and conclusions}\label{sec_summary}
NGC~3521 is a galaxy consisting of two dynamically distinct \hi\ components.  Channel maps and position-velocity slices extracted from the data cube provide clear evidence for an anomalous \hi\ component that is both diffuse and slow-rotating.  The focus of this paper has been to decompose the data cube into its regular and anomalous \hi\ components so that each may be individually modelled and its properties quantified.  

Parameterising the \hi\ line profiles in the cube as a two-component Gaussian allowed for a set of simple criteria to be used to identify the regular and anomalous \hi\ components.  The mass of all  the anomalous \hi\ in the galaxy is estimated to be  $\sim 1.5\times 10^9$~\msun\ - approximately 20~per~cent of the total \hi\ mass.  This is similar to the estimated masses of anomalous \hi\ components seen in other nearby galaxies.  A set of standard \hi\ data products have been produced for each of the regular and anomalous mass components.  Throughout the galaxy, the \hi\ total intensity maps show the anomalous \hi\ to typically constitute less than $\sim 30$~per~cent of the total flux in a line profile.  The velocity fields of both \hi\ components are indicative of circular rotation.  

Tilted ring models were fitted to the velocity fields in order to derive a rotation curve for each \hi\ component.  The results strongly suggest the anomalous \hi\ to be lagging the rotation speed of the regular \hi\ by 25~-~100~\kms\ throughout most of the galaxy.  The tilted ring models were converted into three-dimensional models of the \hi\ data cubes in order to study the distribution of the regular and anomalous \hi.  While thin-disc models for the regular \hi\ are able to match the observations, the anomalous \hi\ clearly needs to be distributed in a much thicker layer.  Models with a disc scale-height of $\sim 3$~-~4~kpc for  the anomalous \hi\ are able to reproduce many of the main features seen in position-velocity slices.  

The anomalous \hi\ in NGC~3521 is most likely a slow-rotating mass component residing in the halo of the galaxy.  The various results presented in this work are very similar to those presented by \citet{oosterloo_NGC891} and \citet{Swaters_NGC891} who infer the presence of a galactic fountain in the edge-on galaxy NGC~891.  {\color{black}The distribution of  anomalous \hi\ in NGC~3521 is shown to be spatially coincident with the inner regions of the disc where the star formation rate is highest.  This provides strong evidence for a link between stellar feedback and the anomalous halo gas.} Some of the halo gas is also expected to be accreted from the intergalactic medium (IGM), {\color{black}yet no direct evidence is provided  in this work for the presence of material originating from outside of the galaxy system.}

%ACKNOWLEDGEMENTS----------------------------------------------------------------------------------------------------------------------------------------------------------------------------------------------
% !TEX root = main.tex
\section{Acknowledgements}
This research has been supported in part by the South African Research Chairs Initiative (SARChI) of the Department of Science and Technology (DST), and the National Research Foundation (NRF).  This publication makes use of data products from the Wide-field Infrared Survey Explorer, which is a joint project of the University of California, Los Angeles, and the Jet Propulsion Laboratory/California Institute of Technology, funded by the National Aeronautics and Space Administration.  This research has made use of the NASA/IPAC Extragalactic Database (NED) which is operated by the Jet Propulsion Laboratory, California Institute of Technology, under contract with the National Aeronautics and Space Administration.  {\color{black}Sincere thanks are extended to the anonymous referee for providing insightful, very useful feedback that improved the overall quality of the paper.  }

\bibliographystyle{mn2e} 
%\bibliography{/Users/ed/latex/bibliography}{}

\label{lastpage}

\end{document}